\documentclass{nature_pic}

\usepackage{xcolor}

\title{Photon quantum entanglement in the MeV regime and its application in PET imaging}

\author{D.P. Watts$^{1}$*, J. Bordes$^{1}$, J.R. Brown$^{1}$, A. Cherlin$^2$, R. Newton$^{1}$, J. Allison$^{3,4}$, M. Bashkanov$^{1}$, N. Efthimiou$^{1,5}$, N.A. Zachariou$^{1}$ }

\usepackage{graphicx}
\usepackage{mathtools}
\usepackage{braket}
\usepackage{subcaption}
\usepackage{lineno}

\begin{document}

\maketitle

\begin{affiliations}
\item Department of Physics, University of York, Heslington, York, YO10 5DD, UK.
\item Kromek Group plc, Sedgefield, County Durham, TS21 3FD, UK.
\item Geant4 Associates International Ltd., Hebden Bridge, HX7 7BT, UK.
\item Department of Physics and Astronomy, University of Manchester, Manchester, M13 9PL, UK.
\item PET Research Centre, School of Health Sciences, University of Hull, Hull, HU6 7RX, UK.
\end{affiliations}

\begin{abstract}
Positron Emission Tomography (PET) is a widely-used imaging modality for medical research and clinical diagnosis. Here we demonstrate, through detailed experiments and simulations, an exploration of the benefits of exploiting the quantum entanglement of linear polarisation between the two positron annihilation photons utilised in PET. A new simulation, which includes the predicted influence of quantum entanglement on the interaction of MeV-scale photons with matter, is validated by comparison with experimental data from a cadmium zinc telluride (CZT) PET demonstrator apparatus. In addition, a modified setup enabled the first experimental constraint on entanglement loss for photons in the MeV regime. Quantum-entangled PET offers new methodologies to address key challenges in next generation imaging. As an indication of the potential benefits, we present a simple method to quantify and remove in-patient scatter and random backgrounds using only the quantum entanglement information in the PET events.
\end{abstract}

\section{Introduction}

Positron Emission Tomography (PET) is a valuable imaging tool for exploring a variety of cellular and molecular processes \textit{in vivo}. PET can provide high contrast and quantitative functional information about disease or therapy response usually complemented by purely anatomical information provided by other imaging modalities. A typical PET study involves the administration of a radiotracer, a biologically active molecule which is labelled with a positron ($e^{+}$) emitting radionuclide to track metabolic activity. Subsequent $e^{+}e^{-}$ annihilation produces two $0.511$ MeV $\gamma$-photons, moving in approximately opposite directions. Their subsequent detection enables a line of response (LOR) to be defined upon which the annihilation site is assumed to be located. However, in addition to this spatial information, the two annihilation $\gamma$ are described by a common entangled wavefunction which results in correlations between their interaction processes even when separated, the so called ``spooky''\cite{EPR,Bohm} action at a distance effect of quantum mechanics. The potential imaging benefits from utilising quantum entanglement information are more extensively investigated for optical ($\sim$eV) photons \cite{Brida2010,Morris2015,Genovese2016}, and more recently for X-ray photons ($\sim$10 keV)\cite{Sofer2019}. 
In this work, we carry out a first simulation of the interaction of MeV scale photons with matter in a PET-like system, which includes the effect of quantum entanglement in their interactions. We explore the benefits of using this quantum information in the analysis of PET data.

The most useful events collected to form a PET image (true events) are those having a LOR which crosses the annihilation site (no interaction in the patient). These true events are expected to maintain their quantum-entangled nature.  However, in PET these are recorded along with unwanted scatter and random backgrounds. These have LORs displaced from the annihilation site(s) which cause artefacts in the reconstructed image\cite{Hoffman1981,scatter}, decrease the signal-to-noise and distort the relationship between the image intensity and the activity in the volume of interest. Scatter background arises when at least one of the two annihilation $\gamma$ scatter prior to detection. In addition to a displaced LOR, such decohering scatter reactions will lead to entanglement loss for the subsequently detected photon pair. The scatter-to-true ratios range from $\sim$0.2 for brain imaging to $\sim$2 for 3D abdominal imaging\cite{scatter_true}. The random background originates mainly from uncorrelated $\gamma$ pairs, producing LORs dispersed over the full image and which would, of course, not be in an entangled state. Random-to-true ratios range from $\sim$0.1 (brain imaging) to more than 1\cite{scatter_true}, influenced by the detector properties (e.g the timing coincidence window for the accepted $\gamma$ photons) and the administered activity.

To investigate the benefits of exploiting quantum entanglement information in PET, a new simulation accounting for the {\em entangled} interactions of annihilation photons in matter was developed. This builds on the comprehensive Geant4 simulation framework enabling full account of detector geometry, experimental resolutions and backgrounds. Therefore, beyond the PET application discussed here, the framework offers possibilities to improve the achievable precision in measurements of entangled photons at the MeV scale, such as precision tests of the predicted DCSc cross sections. To test its validity, the new simulation is compared to high quality experimental data on double Compton scattering (DCSc) of positron annihilation (Pa) gamma from a $^{22}$Na positron source. These new data were obtained using two state-of-the-art semiconductor $\gamma$ detectors which, when placed back-to-back, provided a PET-demonstrator apparatus. Additionally, simulated preclinical PET images from a larger system of such detectors were simulated, to provide assessment through simulation of the benefits of exploiting the new information.

In this first study, we used the linear polarisation of the $\gamma$ as the experimental observable sensitive to the entangled nature of the photons, with visibility achieved through observation of the DCSc process of the photons. The two $\gamma$ from (ground state) para-positronium annihilation (anti-parallel spins; $S=0$, $S_{z}=0$ where $S$ is the spin of the positronium) have orthogonal linear polarisation with an entangled wave function which can be expressed as:
\begin{equation}
    \ket{\Psi} = \frac{1}{\sqrt2}\: \Big( \ket{x}_{-}\ket{y}_{+}  -   \ket{y}_{-}\ket{x}_{+}\Big)
    \label{eqn:wavefunc}
\end{equation}
where $\ket{x}_{-}$ and $\ket{y}_{-}$ represent a $\gamma$ propagating along $-z$ with polarization in $x$ and $y$, respectively. $\ket{x}_{+}$ and $\ket{y}_{+}$ have equivalent definitions but for the $+z$ direction. This entangled Bell state is the only allowed state following annihilation of ground state positronium to two photons\cite{Snyder1948,Thesis,Bohm}.

\begin{figure}
    \centering
    \includegraphics[width=\textwidth]{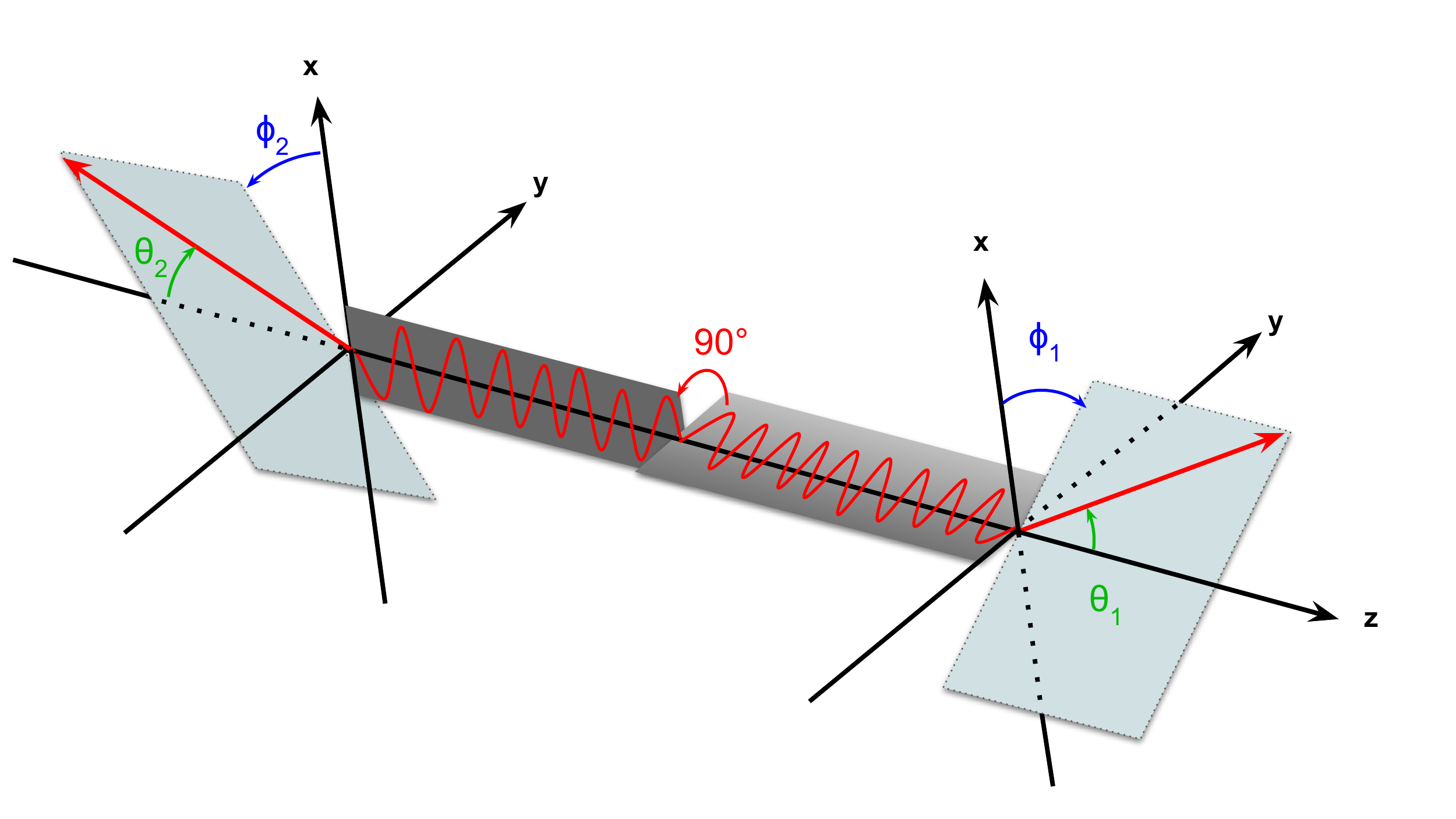}
    \caption{\textbf{Scattering angle definition.} Schematic figure showing the definition of the scattering angles $\theta$ and $\phi$ for the DCSc of the two entangled $\gamma$ photons. The direction vectors for the scattered photons are shown by the red lines, with the corresponding scatter planes indicated by the shaded rectangles. The z-axis is aligned with the $\gamma$ direction while the x-axis is defined with respect to the detector in the laboratory frame. (Arbitrary) mutually perpendicular orientations of the polarisation vectors of the $\gamma$ are also shown.} 
    \label{fig:thetaPhi_Def}
\end{figure}

 The differential cross section for DCSc of the entangled annihilation gamma in the allowed Bell state of equation \ref{eqn:wavefunc}, is given by\cite{Pryce1947,Thesis}: 

\begin{equation}
     \frac{d^{2}\sigma_{double}}{d\Omega_{1}d\Omega_{2}} =\frac{r_{0}^4}{16}\:\Big(K_{a}(\theta_{1},\theta_{2}) - K_{b}(\theta_{1},\theta_{2})\cdot{}\text{cos}(2\Delta\phi)\Big)
     \label{eqn:EKN}
 \end{equation}
where ${d\Omega_{1,2}}$ and $\theta_{1,2}$ are the solid angles and polar scattering angles for $\gamma_{1}$, $\gamma_{2}$, respectively, $r_{0}$ is the classical electron radius, $K_{a}$ and $K_{b}$ are kinematic factors\footnote{$K_{a}=\frac{[(1-cos\theta_{1})^{3}+2]\cdot{}[(1-cos\theta_{2})^{3}+2]}{(2-cos\theta_{1})^{3}\cdot{}(2-cos\theta_{2})^{3}}$, $K_{b}=\frac{sin^{2}\theta_{1}\cdot{}sin^{2}\theta_{2}}{(2-cos\theta_{1})^{2}\cdot{}(2-cos\theta_{2})^{2}}$}, and $\Delta\phi = \phi_{1}-\phi_{2}$ is the relative azimuthal scattering angle (see Fig. \ref{fig:thetaPhi_Def} for definition of scatter angles). The form of equation 2 is presented in the publication of Pryce and Ward\cite{Pryce1947} (also independently derived by Snyder \textit{et al.}\cite{Snyder1948}). In these early works, consistent results were obtained when employing time dependent perturbation theory or a Klein-Nishina\cite{kleinNishina} based approach, as outlined  in the PhD thesis of Ward\cite{Thesis}. Subsequently the same form has been derived in DCSc calculations in a matrix formalism\cite{Caradonna2019} and employing Kraus operators\cite{Hiesmayr2019}.

 The DCSc cross section is modulated by the cos$(2\Delta\phi)$ term, resulting in an enhancement ratio ($R$) between the maximum ($\Delta\phi=\pm90^\circ$) and minimum ($\Delta\phi=0^\circ$) scattering probabilities of $R=2.85$, achieved when both $\theta_{1}$ and $\theta_{2}$ are equal to $81.7^\circ$ \cite{Snyder1948,Pryce1947,Thesis}. Bohm and Aharonov\cite{Bohm} were the first to recognize that the  ($\Delta\phi$) correlations between the DCSc annihilation photons was an example of the kind of entanglement discussed by Einstein, Podolsky and Rosen\cite{Einstein}. They also derived\cite{Bohm} the upper limit for a (hypothetical) orthogonally polarised, but non-entangled (separable) state as $R=1.63$, establishing that measured values above this limit are a witnesses of the entanglement\cite{Bohm, Hiesmayr2019, Caradonna2019}. Experimental measurements of this amplitude \cite{Wu1950,Langhoff1960,Kasday1971,Faraci1974,Kasday1975,Wilson1976,Bruno77,Bertolini1981} focused on restricted kinematics of $\theta_{1}$, $\theta_{2}$ around maximum visibility, and yielded $R$ values  well beyond the upper limit for a non-entangled state. Clear statistical agreement with the entangled theory (equation 2) was established (with analytic corrections for experimental aspects of the measurement).\footnote{The most precise measurements were obtained by Langhoff\cite{Langhoff1960} and Kasday \textit{et al.}\cite{Kasday1975} with measurements  $R=2.47\pm0.07$ and $R=2.33\pm0.10$ respectively, in agreement with equation 2 when geometrical effects are accounted for.}
 The consistency of the measured $R$ was established\cite{Bruno77,Wilson1976} for separation distances up to $2.5$ m, greatly exceeding the coherence length associated with Pa ($0.12 $ m\cite{Wilson1976}) and larger than the diameter of a typical clinical PET scanner (0.8-1.3~m).
 The DCSc entanglement witness was also exploited in early investigations of Bell's inequalities, providing evidence against (although not fully ruling out) hidden-variable theories in quantum mechanics\cite{Bertolini1981,Bruno77,Kasday1975}. Subsequently, optical EPR measurements proved more amenable due to the availability of more efficient linear polarisation analysers, which enabled tests with less auxiliary assumptions\cite{Clauser_1978}. However, we remark that utilisation of the (experimentally verified) correlation in DCSc scatter planes in this work (equation 2), does not rely on the existence of loop-hole free EPR tests. The ability of the entangled cross section (equation 2) to describe the $\Delta\phi$ distribution ($R$) for all previous data in positron annihilation (and for the current data - see next section) gives confidence in its implementation in PET simulation.\footnote{The small contribution of annihilation in flight, in which the positron annihilates before thermalisation, is neglected as assumed for previous Pa measurements\cite{Wu1950,Langhoff1960,Kasday1975,Wilson1976,Bruno77,Bertolini1981,Bohm} and only contributes  in tissue at the 2\% level e.g\cite{Harpen04}. We neglect three-$\gamma$ decays from ortho-positronium ($S=1$) as they have a yield only 0.5\% of the two-$\gamma$ yield in PET\cite{Harpen04}, and could be further rejected by photon energy cuts. Initial  ortho-positronium formation may lead to two-$\gamma$ final states through singlet ($S=0$) pickup annihilation with an (external)  electron in the surrounding medium e.g.\cite{Shibuya20}. We take such photons as having the correlated wavefunction of equation 1 (and DCSc cross section of equation 2) as implicit in previous measurements. This assumption is supported by the observation that, despite a range of positron thermalising media being employed, no deviation from equation 2 is observed in any previous experimental work on Pa\cite{Wu1950,Langhoff1960,Kasday1975,Wilson1976,Bruno77,Bertolini1981,Bohm} (or the current work - see next section).}
 

 Establishing the utility of quantum entangled PET (QE-PET) required the  implementation of the predicted effects of quantum entanglement into a simulation of PET-photon interactions in matter. We incorporated entanglement (as described by equation \ref{eqn:EKN}) in the leading Monte Carlo simulation package, Geant4\cite{Agostinelli2003,Allison2016}, by developing new routines which enable communication between the individual particle-tracking processes to reproduce the scattering cross sections for entangled $\gamma$ pairs (equation 2). The new version, which we refer to as {\em QE-Geant4}, will be included in future releases of the code. Any further interactions of $\gamma$ after the initial Compton scatters reverted to the standard Geant4 polarised Compton scattering routines, \textit{i.e.} a Compton interaction was assumed to completely collapse the entangled state shown in equation 1.

 The QE-Geant4 simulation was validated by comparing to measurements of DCSc of Pa photons in a PET-demonstrator. 
 The system was developed by Kromek Group based on the DMatrix detector system\cite{McAreavey2017} and comprised two 10~mm cubic semiconducting cadmium zinc telluride (CZT) crystals, placed back-to-back and separated by 55~mm. A segmented anode divided each crystal into 121 $0.8 \times 0.8$~mm$^{2}$ pixels, with depth information accessible from the anode drift time\cite{Zhang2005}. Readout of the anode and cathode signals employed a bespoke ASIC. The coincidence timing window of the system was set to 1~$\mu$s due to the charge sweep-out time of a few tenths of $\mu$s. A 170~kBq $^{22}$Na source, comprised of a 1mm diameter active bead housed in a thin plastic was placed equidistant from each crystal face providing a source of annihilation $\gamma$. 

\begin{figure}
\centering
\includegraphics[width=0.8\textwidth]{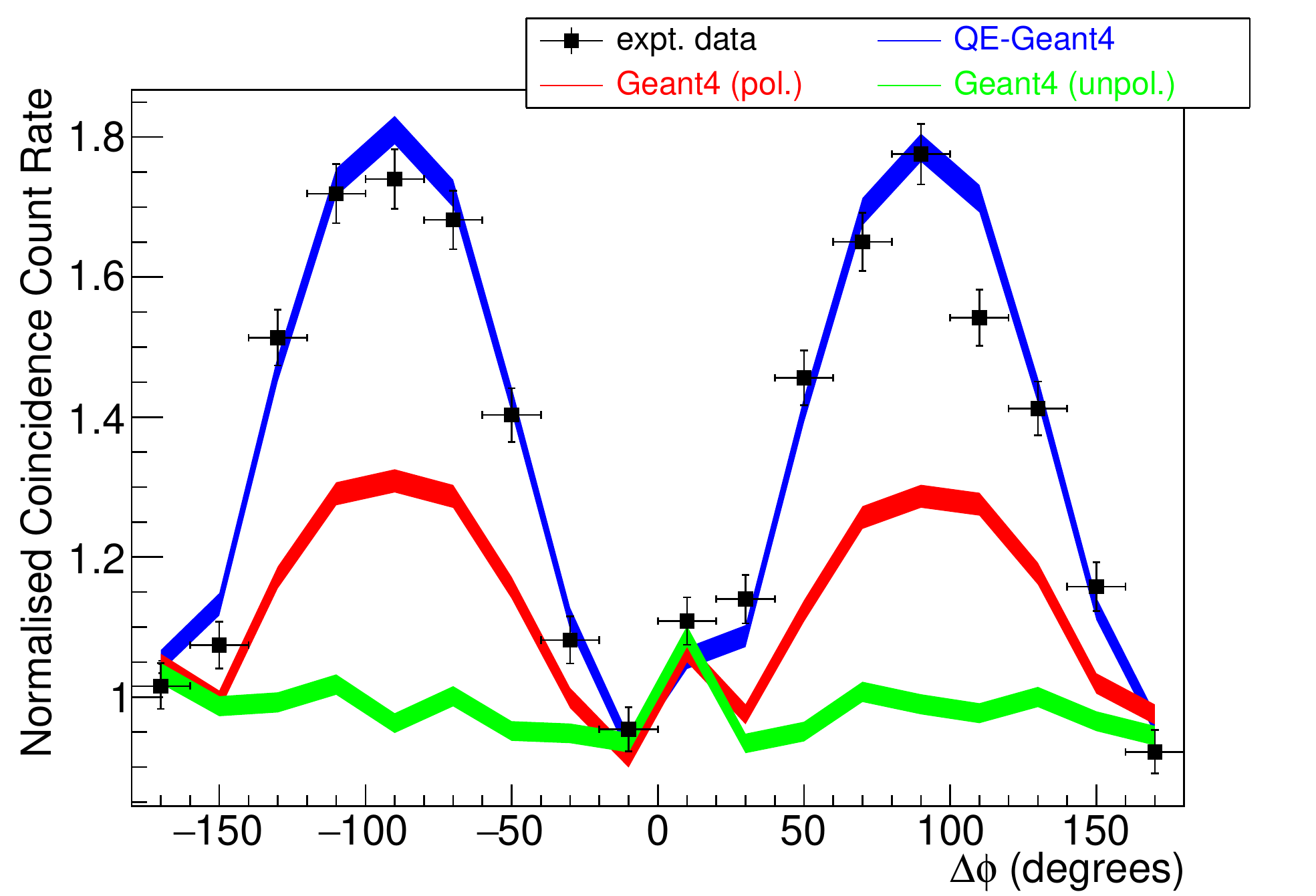}
\caption{\label{fig:data}\textbf{Comparison of experimental and simulated scattering probabilities.} Black data points show the experimentally determined (normalised) coincidence count rate as a function of $\Delta\phi$, obtained using the CZT detector apparatus (see text). The prediction from the quantum-entangled simulation (QE-Geant4) is shown as the blue line. Non-entangled predictions for orthogonally polarised $\gamma$ pairs are shown as the red line. The prediction for unpolarised annihilation $\gamma$ is shown by the green line. The statistical uncertainties for the simulations are indicated by the line widths. For comparison of the $\Delta\phi$ distributions, the experimental and simulated data are normalised to unity for the data around the minima at $\Delta\phi=0^{\circ}$ and $\pm180^{\circ}$, specifically the mean yields from bins 1, 9, 10 and 18.}
\end{figure}

Measurements of $\Delta\phi$ were accessible for events where both annihilation $\gamma$ interacted through a Compton scattering and a subsequent photoelectric absorption. Such events produced, in both CZT modules, two clusters of pixels with a total energy in the range 480 - 530 keV. Charge sharing events, produced by a single hit between adjacent pixels, were rejected by requiring a gap of at least one pixel between clusters. The polar scattering angle was determined using the Compton scattering formula, assuming the largest energy signal as the first interaction\footnote{Simulations indicate that this assignment will be correct in $\sim60\%$ of cases. Incorrect assignment will result in a calculated polar scattering angle $\theta_{calc.}=180^\circ - \theta_{real}$, but not affecting $\Delta\phi$ due to the symmetry about $0^\circ$.}. To enhance the amplitude of the cos($2\Delta\phi)$ distribution (equation \ref{eqn:EKN}), coincidences were only retained if the polar scattering angle was within $70^\circ{}\le\theta\le110^\circ{}$. The azimuthal scattering angle was simply determined from the energy-weighted centre-of-gravity of each cluster and only one- or two-pixel clusters were considered. The azimuthal angular resolution depended on the distance between the clusters. It ranged from $2.9^\circ$ (for the most distant pixels) to $20.4^\circ$ for the closest pixels considered in this analysis. The energy and $\theta$ cuts reduce the data to 29.2\% and 5.4\% of the total two-cluster events respectively, and when combined retain 2.6\% of the total yield. The normalised (see caption) coincidence count rate as a function of $\Delta\phi$ for these events is shown as the data points in Fig. \ref{fig:data}.
A strong cos($2\Delta\phi)$ modulation is seen with a {\em measured} enhancement factor of $R=1.85 \pm 0.04$ for the event yield selected by the analysis cuts. 
\footnote{Selecting a tighter polar angle acceptance ($\theta_{1,2}=93-103^{\circ}$) gave a {\em measured} enhancement factor of $R=1.95\pm0.07$. We note that future higher statistics data could further increase the visible enhancement by selecting sub-set of events with larger intercluster separations and correspondingly improved angular resolution.}
 
The experimental setup was simulated with QE-Geant4. The simulated energy deposits were matched to the CZT experimental detector response by accounting for effects of charge transport, diffusion and self-repulsion\cite{Knoll2010}, and smeared to match the experimental energy resolution of each detector module (3.8 and 5.3\% FWHM at 662 keV). The resulting simulated data were then passed through the same analysis code as the experimental data. 
 
The predictions from QE-Geant4 (Fig.~\ref{fig:data} - blue line) show a good agreement with the measured $\Delta\phi$ distribution ($\chi^2/\nu=1.87$), providing validation of QE-Geant4 simulation as well as the underlying entanglement theory employed (the agreement between the measured $\Delta\phi$ distribution and QE-Geant4 is also quantified on a bin-by-bin basis in the supplementary materials). To better quantify the effects of entanglement on the measured $\Delta\phi$ amplitude, a further simulation of (hypothetical) non-entangled, but orthogonally polarised, annihilation $\gamma$ is also presented (red line). The DCSc for this case is modelled by the standard (non-entangled) polarised Compton scattering classes in Geant4, which model the scatter of each $\gamma$ as an independent (separable) process \footnote{As shown in the supplementary materials the simulation results for both the entangled and non-entangled systems agree with the analytical DCSc formulae derived for each hypothesis by Bohm and Aharonov\cite{Bohm}.}.
The size of the predicted $\Delta\phi$ modulation is clearly reduced if entanglement is neglected, and the predictions are in clear disagreement with the experimental data. The results illustrate the need for an account of quantum entanglement in PET simulation to correctly describe the observed distribution of Compton scatter planes\footnote{Earlier works simulating (non-entangled) polarisation effects in PET using standard Geant4 classes\cite{McNamara:2014,Toghyani2016} could be further developed using this new entangled framework.}. In order to assess the influence of detector acceptance on the measured $\Delta\phi$ distribution, a further simulation for unpolarised annihilation $\gamma$ was also carried out (green line). The intrinsic $R$ for such events is equal to unity, in agreement with analytical calculations for mixed states in Bohm and Aharonov\cite{Bohm} and Caradonna \textit{et al.}\cite{Caradonna2019}.\footnote{We note that a very recent work\cite{Hiesmayr2019} contradicted the original theory of Bohm and Aharonov\cite{Bohm} and the interpretation of all previous experimental works\cite{Bruno77,Wilson1976,Bohm,Snyder1948,Pryce1947,Bertolini1981,Kasday1975,Kasday1971,Langhoff1960,Faraci1974,Wu1950}, by inferring a contribution to $R$ from a hypothetical mixed state in Pa (in which the $\gamma$ need not originate from the same annihilation event). The conclusion has already been refuted by Caradonna \textit{et al.}\cite{Caradonna2019} who re-derived $R=1$ for mixed states in a matrix formalism in agreement with Bohm and Aharonov\cite{Bohm}. We note the unpolarised simulation presented here also supports that such events are consistent with $R=1$. We therefore adopt the accepted theoretical interpretation\cite{Bohm} in this work.}
The $\Delta\phi$ distribution for unpolarised events in the detector acceptance (green line) is rather uniform albeit with a small acceptance related enhancement near to the centre of the distribution, which is also evidenced in the experimental data, and appears well modelled by the simulation. It is clear that the measured $\Delta\phi$ distribution only has modest influence from detector acceptance effects.


\subsection{Investigation of entanglement loss when an annihilation gamma scatters prior to measurement}

The results in Fig 2, and previous measurements in more limited kinematics from a range of different Pa sources, show that the azimuthal correlation of the Compton scatter planes in DCSc of Pa photons is in agreement with the entangled theory (equation 2) and has a correlation ($R$) beyond the upper limit of a separable non-entangled state. We remark that the data in Fig 2 is dominated by photons for which their first interaction is the (linear polarisation analysing) Compton scatter reaction in the CZT detectors. We are not aware of any previous measurement of the $\Delta\phi$ correlation for photons which have undergone an identified decohering process {\em prior} to the measurement. 

\begin{figure}
\centering
\includegraphics[width=1.0\textwidth]{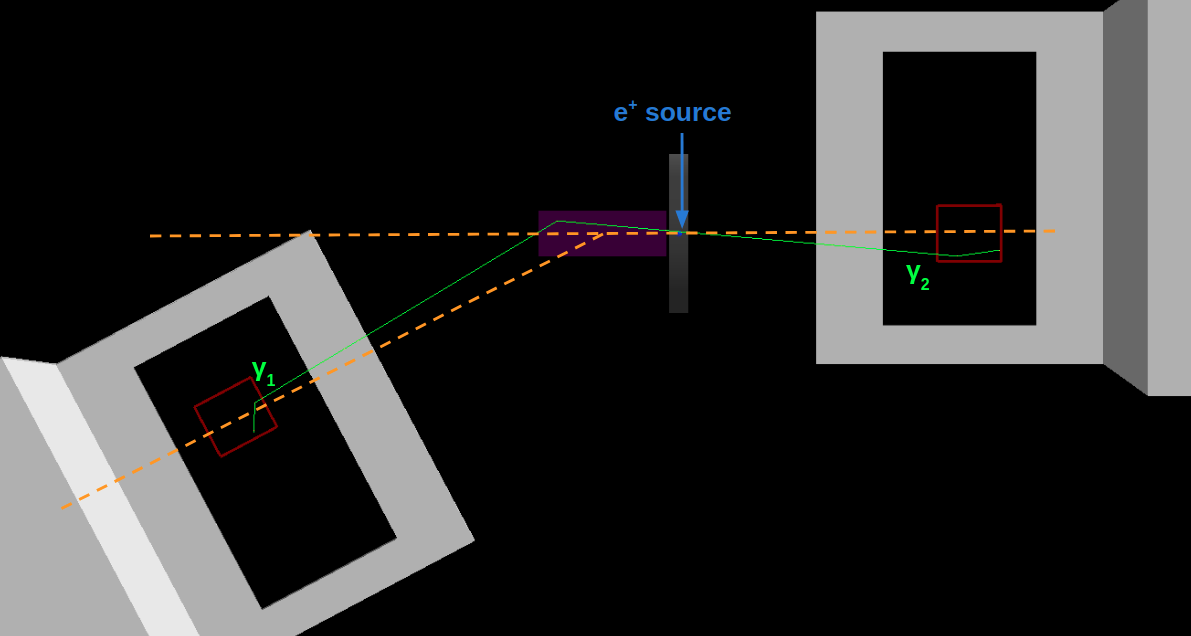}
\caption{\label{fig:scatG4}\textbf{Geant4 visualisation of the experimental setup for the scattering measurement.} The CZT crystals (red) are shown along with their support structures (grey) and the nylon scattering medium (purple). The topology of a typical scatter event are shown by the solid lines. The initial back-to-back trajectory of the two annihilation $\gamma$ from a single Pa event can be seen as the green lines originating at the source position. One of the photons Compton scatters in the nylon scattering medium (purple). The subsequent Compton scatters of both $\gamma$ within the CZT crystals, from which the $\Delta\phi$ correlation is obtained, are evident from the kinks in the photon trajectories.}
\end{figure}

 To achieve this we adapted the experimental setup  as shown in Fig.~\ref{fig:scatG4}. Events where one of the photons has undergone a Compton scattering process prior to measurement of the $\Delta\phi$ correlation ( {\em prior-scatter} events) were obtained by inclusion of a scattering medium (nylon) in the path of one of the annihilation $\gamma$, with the corresponding CZT module rotated through 33$^{\circ}$ relative to the centre of the scatterer. The measured energy of the scattered $\gamma$ in the CZT (obtained from the sum of the two energy deposits) matched that expected from the reaction kinematics ($\sim$440~keV) and was well separated from backgrounds.  
 
\begin{figure}
\centering
\includegraphics[width=0.8\textwidth]{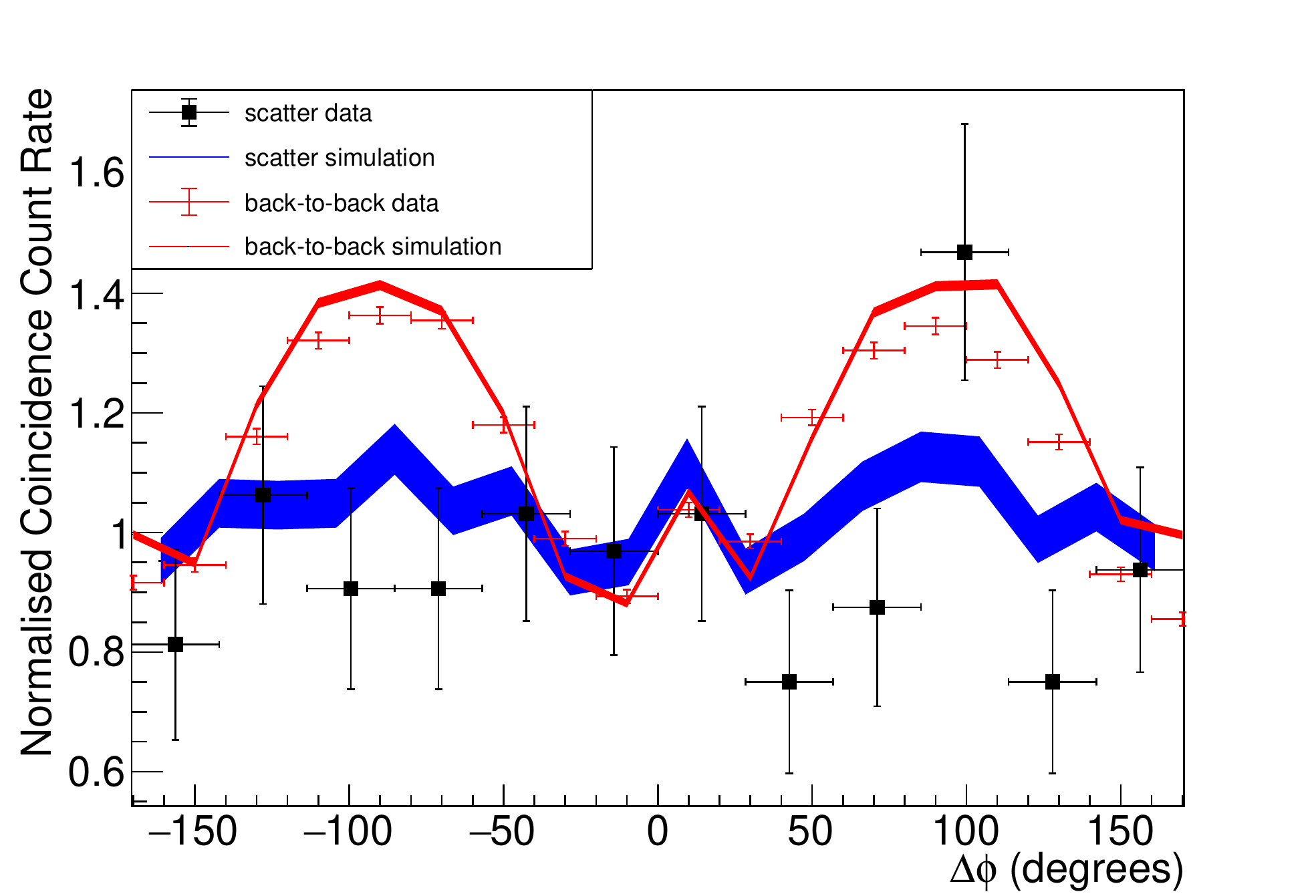}
\caption{\label{fig:scatData}\textbf{$\Delta\phi$ distributions for events with a {\em prior-scatter} process} The black data points were obtained with the setup of Fig 3 and show the measured $\Delta\phi$ distribution when one of the entangled $\gamma$ has scattered through a polar angle of $\sim$33$^{\circ}$ prior to the detection in the CZT. QE-Geant4 predictions for this setup are shown as the blue line. The red data points show the measured $\Delta\phi$ distribution without scatterer, in a back-to-back configuration,   for the same polar scatter angle range. The red line shows the QE-Geant4 predictions corresponding to this latter setup. All simulations and data employ a selection of polar scatter angles (within the CZT) in the range $60^\circ\leq{}\theta\leq{}140^\circ$.
}
\end{figure}

 The DCSc $\Delta\phi$ distributions measured by the CZT detectors for such {\em prior-scatter} events are shown by the black data points in Fig.~\ref{fig:scatData}. For comparison, the red data points (red line) show the measured (simulated) $\Delta\phi$ correlations for back-to-back unscattered $\gamma$ respectively, with the same polar angle cuts  for the scattered photons as used for the {\em prior-scatter} data\footnote{To increase the event yield both data sets have a wider polar angle acceptance for the CZT scatters than employed for Fig.~\ref{fig:data}. (see caption)}. The measured $\Delta\phi$ correlation for the {\em prior-scatter} events is clearly diminished compared to the unscattered case. The QE-Geant4 prediction (blue line) also exhibits a diminished $\Delta\phi$ correlation, in statistical agreement with the experimental data. In the QE-Geant4 modelling a complete loss of entanglement is assumed following the first DCSc (predominantly in nylon and CZT for this event sample). Any subsequent gamma interactions are modelled as for polarised, independent $\gamma$ (i.e. a separable state). 
 
 The consistency between the {\em prior-scatter} data and the corresponding simulation predictions indicates the assumption of wavefunction collapse between Pa $\gamma$ following a scatter process
 is consistent with the data. The magnitude of the $\Delta\phi$ correlation for {\em prior-scatter} events is also clearly diminished compared to the unscattered case. As far as we are aware the world's current data on this process is contained in Fig 4, so clearly a future measurement programme with improved statistical accuracy and a wider range of scattering kinematics are a clear next step for the field. Such measurements (although planned) are currently out of reach of the small demonstrator system used here\footnote{The scatter data in Fig.~\ref{fig:scatData} required over a month of data taking}.

\subsection{Quantum entangled PET}

\begin{figure}
\centering
\includegraphics[width=0.5\textwidth]{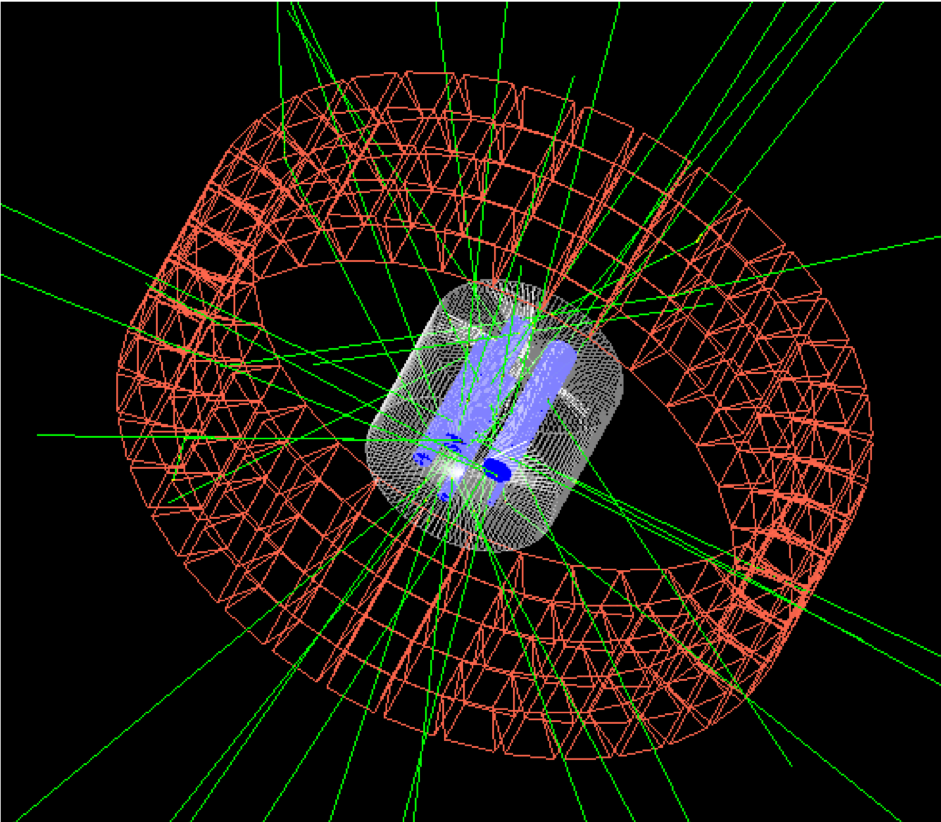}
\caption{\textbf{Geant4 visualisation of the simulated preclinical PET scanner.} The scanner consisted of four rings of 10~mm cubic CZT detector crystals (orange). The NEMA-NU4 phantom consisted of a cylinder of PMMA (white) and five rods which were filled with a solution of liquid water and $e^{+}$ source (blue). The simulated trajectories of 20 pairs of annihilation $\gamma$ are shown in green.}
\label{fig:scanner}
\end{figure}

 The ability of the new QE-Geant4 to accurately describe the observed $\Delta\phi$ correlations for DCSc of Pa photons offers new possibilities to separate the true (entangled) PET events from the backgrounds arising from scatter and random coincidences, exploiting the differences in the $\Delta\phi$ correlations to identify their contribution to the image on a statistical basis (note the correlations do not offer the possibility to identify contributions on an event-by-event basis). To illustrate this potential, PET images were produced from a QE-Geant4 simulation of a preclinical PET acquisition. A scanner composed of four rings of CZT (with the same pixel size as the demonstrator) was defined, along with a standard preclinical mouse phantom NEMA-NU4 (National Electrical Manufacturers Association)\cite{nema} (Fig. \ref{fig:scanner}). The phantom consisted of a cylinder of tissue equivalent poly(methyl methacrylate) (PMMA) with five capillaries (1-5 mm in diameter) filled with a solution of water and an $e^{+}$ source.
  
As a first study of QE-PET imaging benefits, two-dimensional PET images were reconstructed from the simulated data using a simple filtered back-projection (FBP) algorithm\cite{Ramachandran1971} and are shown in Figs. \ref{fig:scatter}a and \ref{fig:random}a, corresponding to simulations with scatter or random backgrounds respectively. Only coincidences where both $\gamma$ have polar scattering angles in the CZT within $67^\circ\le\theta\le97^\circ$ were retained, as a compromise between keeping the enhancement ratio high and maintaining good statistics. The FBP images clearly exhibit the structure of the NEMA-NU4 phantom, showing activity from the 5 capillaries. The intensity profiles for a region of interest crossing the two lowest capillaries (seen in yellow in Figs. \ref{fig:scatter}a and \ref{fig:random}a) were extracted for two bins of $|\Delta\phi|$, $0^\circ\le|\Delta\phi|\le20^\circ$ and $80^\circ\le|\Delta\phi|\le100^\circ$. The bins correspond (respectively) to regions where low and high fractions of true events are expected (cf. Fig. \ref{fig:data}). The profile for the true events can be obtained using a simple subtraction of the profiles from each $|\Delta\phi|$ bin, scaled with factors obtained from the QE-Geant4 simulation (see supplementary materials). The results of this subtraction for data sets with scatter (random) backgrounds are shown by the red lines in Fig. \ref{fig:scatter}d (Fig. \ref{fig:random}d). Good agreement is observed with the ``actual'' true profiles (blue lines), which are reconstructed exclusively from the true coincidences identified using \emph{a piori} knowledge from the simulations.  An equivalent methodology can be used to extract the scatter and random distributions in isolation (Figs. \ref{fig:scatter}c and \ref{fig:random}c). The overall magnitude and shape of the QE-PET and ``actual'' profiles are in good agreement for both scatter and random background scenarios. The fluctuations in the extracted profile (more prominent in Fig. \ref{fig:scatter}c) are not statistical in nature and appear to be caused by artefacts in the FBP imaging produced by incomplete acceptance of the array\cite{Hallen2020}~\footnote{The general features of these fluctuations are mirrored in the (statistically independent) profiles from each $\Delta\phi$ bin and are therefore not dominated by statistical effects. Such fluctuations have been observed previously in FBP\cite{Hallen2020} and post-processing methodologies have been attempted\cite{Edhoim1986,Karp1988,Buchert1999}. Note that for these initial assessments, additional processing of the FBP image was not applied.}. In each case, a $4^{th}$ order polynomial fit to the profile is additionally included to enable the average trend to be compared with the ``actual" distribution.


Even this simple illustrative method, which only uses a fraction of the available data from two of the $\Delta\phi$ bins, along with information from the entangled simulation, indicates the feasibility of quantitative assessment of the shape and magnitude of the image backgrounds with QE-PET\footnote{The previous simulation studies of (non-entangled) polarised PET\cite{Toghyani2016} can be improved further using a quantum entangled simulation as described here.}. 

\section{Discussion}

\begin{figure}
    \centering
     \includegraphics[width=0.95\textwidth]{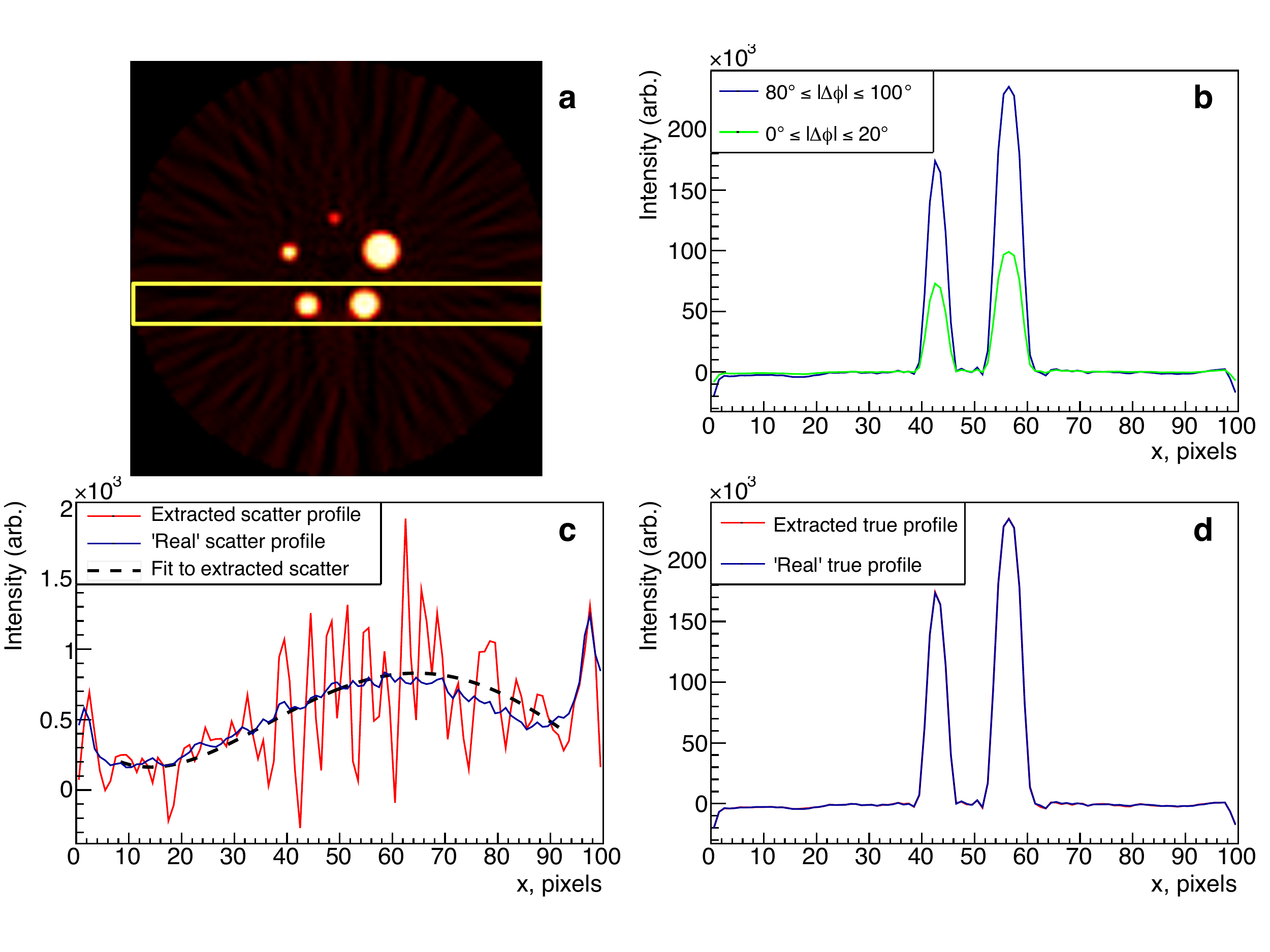}
    \caption{\textbf{Extraction of true and scatter contributions.} (\textbf{a}) FBP two-dimensional PET image of the NEMA-NU4 phantom for true events with a scatter background. (\textbf{b}) Intensity profiles through the region indicated by the yellow rectangle for different $\Delta\phi$ cuts, i.e. $0^\circ\le|\Delta\phi|\le20^\circ$ (green), and $80^\circ\le|\Delta\phi|\le100^\circ$ (blue). (\textbf{c}) QE-PET profile for scatter background events extracted from a scaled subtraction of the two $\Delta\phi$ cut profiles (red line) (see text). The blue line shows the profile from the ``actual'' scatter events in isolation using \emph{a priori} information from the simulation. The dashed line is a $4^{th}$ order polynomial fit to the extracted scatter profile. (\textbf{d}) Profile extracted for true events with QE-PET (red line) compared to the profile of ``actual'' true events in (blue line).}
    \label{fig:scatter}
\end{figure}

\begin{figure}
    \centering
     \includegraphics[width=0.95\textwidth]{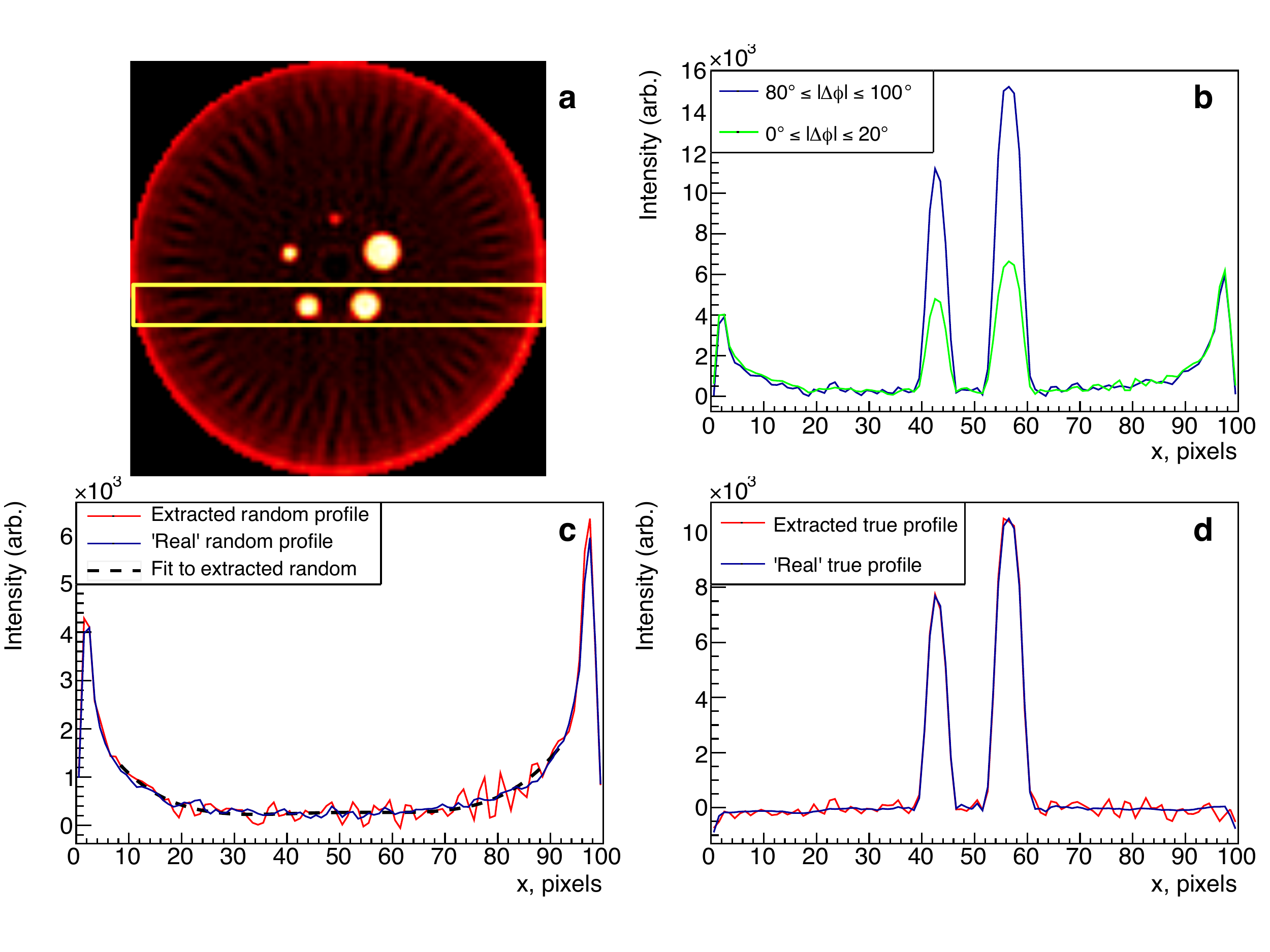}
    \caption{\textbf{Extraction of true and random contributions.} As for Fig. \ref{fig:scatter} but for true events with a random background.}
    \label{fig:random}
\end{figure}

We should remark that the information from QE-PET, as illustrated above, would be obtained in addition to the single-pixel yield routinely analysed to produce PET images. The QE-PET derived information on the backgrounds would equally well apply to these standard PET events. The results (Fig. \ref{fig:scatter} and Fig. \ref{fig:random}) were obtained from a simulation of 10$^{12}$ $e^{+}e^{-}$ annihilations, which is of the same order of magnitude as the radiotracer cumulated activity in a patient during a typical clinical PET scan (a few 100 MBq activity integrated over an acquisition time of 30 minutes). The accuracy in the extracted background profiles are therefore indicative of what may be achieved in a PET scan for this simulated detector geometry\footnote{In a real PET scan scenario, a combination of random and scatter backgrounds are present in the data. In applying QE-PET, their combined contribution would be extracted simultaneously. However, the random coincidence rate can be determined with the delayed window method\cite{Brasse2005}, hence the contributions of scatter and random coincidences could be separated if required.}. We should also note that these first proof-of-principle results employed restrictive $\theta$ and $\Delta\phi$ cuts, thereby only using a fraction of the available data. In future work, these cuts will be optimised to further improve the achievable accuracy by accounting for the interplay between enhancement magnitude and event yield, as explored previously\cite{Toghyani2016,Moskal2018}. However, ultimately we view the optimal use of the new information would be within the framework of more sophisticated imaging methodologies such as the forward model of a MLEM (maximum-likelihood expectation-maximization)\cite{Shepp1982} image reconstruction algorithm. Currently such approaches model scatter-backgrounds employing either: scatter simulation algorithms\cite{watson1996,tsoumpas2005ScatterSimulationIncluding}, which require analytical modelling of the scanner; or CPU intensive Monte Carlo methods. Both approaches require detailed anatomical information from a computed tomography (CT) scan and rely on estimates of the underlying activity biodistribution, which is \textit{a priori} unknown and necessitates iterative approaches\cite{Cook2004, scatter, Zaidi2007}. Implementation of the new QE-PET information in such iterative imaging methods\cite{Thielemans2012P} would be an important next step.\footnote{Time-of-flight (TOF) PET methodologies using fast photon detectors have recently been explored to address backgrounds in PET\cite{Van2016}. The QE-PET method can be employed in parallel with TOF information where available. However, for systems where such timing information is unavailable or of insufficient resolution (e.g. semiconductor detectors as studied here), then QE-PET would offer unique opportunities}.

 \section{Conclusion}
 
 We have simulated the predicted effects of quantum entanglement in the interaction of positron annihilation photons with matter, building on the Geant4 simulation framework. The new simulation predictions were validated by comparison with precision experimental data on double Compton scattering of positron annihilation photons, obtained with a CZT pixelated semiconductor PET demonstrator apparatus. The inclusion of quantum entanglement in the simulation for the reaction process was shown to be crucial to correctly describe the measured correlation between the Compton scatter planes. As well as underpinning the QE-PET developments, the new simulation will enable improved accuracy in any future simulations of PET imaging systems in medical imaging, medical research and industrial applications.  
 
 A modified apparatus, incorporating an additional scattering medium,  enabled a first measurement of the Compton scatter plane correlation when one of the (initially entangled) annihilation photons underwent a Compton scatter process {\em prior} to measurement. The experimental data are compatible with simulation predictions which assume the collapse of the entangled wavefunction following the scatter process.
 
 The new simulation was also used to model a CZT based preclinical scanner and obtain a first simulation study of entangled PET imaging. A method to quantify (and remove) spatially-resolved backgrounds from both in-patient scatter and random coincidences is presented, suggesting entangled PET provides new possibilities to address key challenges for next-generation PET imaging.

 \begin{addendum}
 \item [Acknowledgements] We acknowledge input from D. Jenkins to the manuscript. Simulation work was undertaken on the Viking Cluster, which is a high performance computing facility provided by the University of York. We are grateful for computational support from the University of York High Performance Computing service, Viking and the Research Computing team. The work has been supported by funding from Innovate UK EP/P034276/1 and the UK Science and technology Facilities Council (STFC) ST/K002937/1
\end{addendum}
 
 \bibliographystyle{naturemag}
\bibliography{sample}

\section{Supplementary materials}

\subsection{Verification of QE-Geant4 by comparison with analytical theories of quantum entanglement}

\begin{figure}
 \centering
\includegraphics[width=0.8\textwidth]{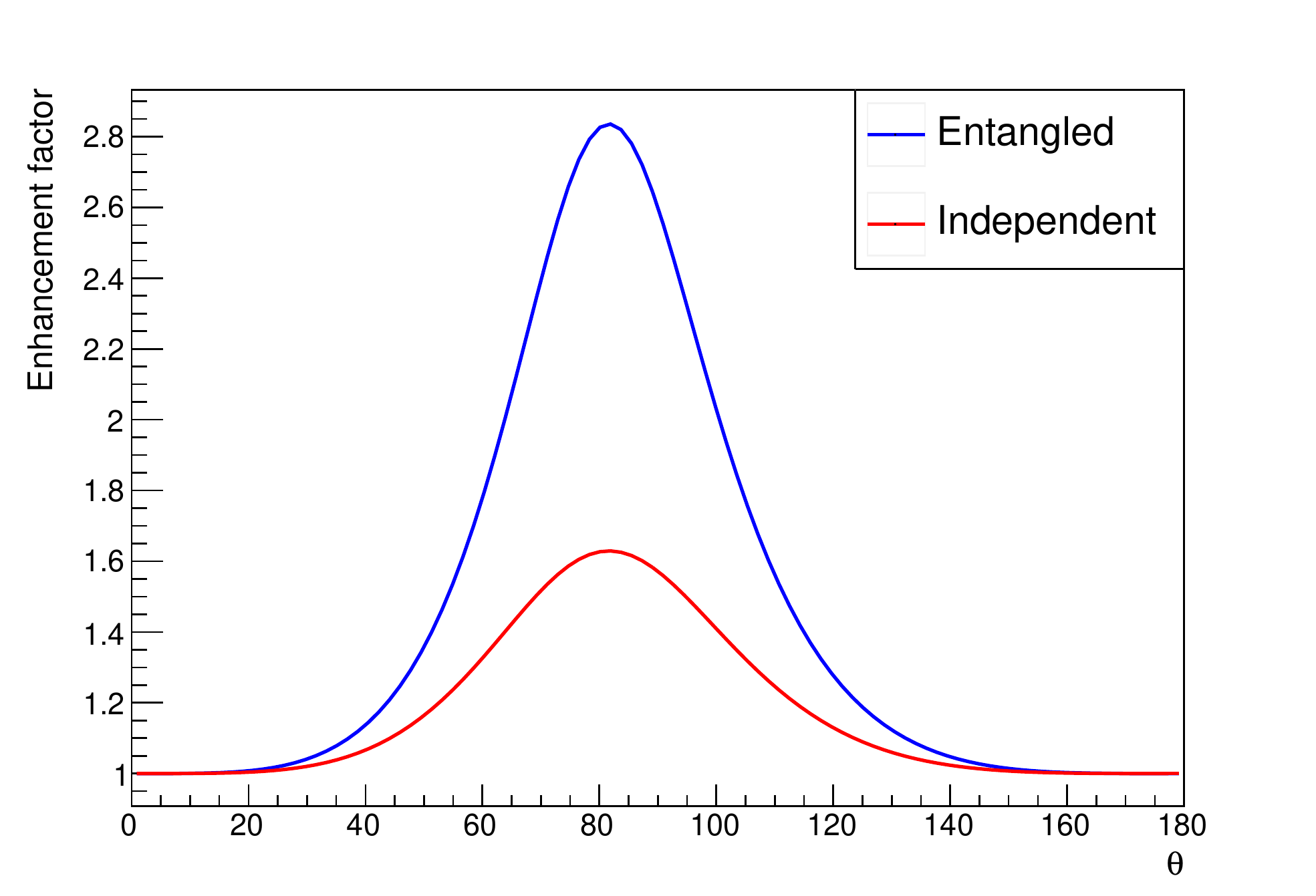}
 \caption{\label{fig:enh}\textbf{Theoretical enhancement factors.} Calculated using the formalism of Bohm and Aharonov\cite{Bohm} for entangled $\gamma$ (blue) and for orthogonally polarised non-entangled $\gamma$ (red).}
\end{figure}

We verified that the simulated effects of quantum-entanglement, implemented in the QE-Geant4 simulation, were in agreement with the theoretical calculations. This was assessed by comparing the normalised coincidence count rate as a function of $\Delta\phi$ for the case of a perfect detector having an effectively infinite detecting medium ($e^{+}$ source placed at the centre of a $10 \times 10 \times 10$~m$^{3}$ cube of CZT) and using the exact $\gamma$ Compton scattering angles from the simulation. In addition, we compared the analytical calculations from a non-entangled hypothesis with the standard Geant4 predictions. The entangled and polarised simulations were carried out with QE-Geant4 and standard Geant4 10.5, respectively. For consistency with the analysis presented in the paper, the same polar scatter angle range for the $\gamma$ was adopted ($67^\circ\le\theta\le97^\circ$). The obtained $\Delta\phi$ distributions for the entangled (non-entangled) scenarios are shown by the blue (red) data points in Fig. \ref{fig:enh}. 

The corresponding theoretical predictions were obtained as follows. From equation \ref{eqn:EKN}, the probability of two entangled $\gamma$ Compton scattering with a relative azimuthal angle $\Delta\phi$ takes the form:
\begin{equation}
    P(\theta_1,\theta_2) = A(\theta_1,\theta_2)\cdot{}cos(2\Delta\phi) + B(\theta_1,\theta_1).
    \label{eqn:cos2phi}
\end{equation}
For clarity, we focus on $\theta_1=\theta_2=\theta$ annihilations and define the enhancement factor ($Enh$) as the ratio of the perpendicular ($\Delta\phi=\pm90^\circ$) and parallel ($\Delta\phi=0^\circ$ and $\Delta\phi=\pm180^\circ$) scattering probabilities:
\begin{equation}
    Enh(\theta) = \frac{P_{\perp}(\theta)}{P_{\parallel}(\theta)}.
    \label{eqn:enh}
\end{equation}
In table 1 (A) of Bohm and Aharonov\cite{Bohm}, the parallel and perpendicular scattering probabilities for entangled $\gamma$ are given by:
\begin{equation}
    P_\parallel(\theta) = 2\gamma(\gamma-2sin^2\theta) 
    \quad\mathrm{and}\quad 
    P_\perp(\theta) = (\gamma-2sin^2\theta)^2+\gamma^2
\end{equation}
where $\gamma = (k_0/k) + (k/k_0)$, with $k_0$ and $k$ the wave numbers of the incident and scattered photons, respectively. The resultant enhancement factor is plotted as the blue line in Fig. \ref{fig:enh}. Note that Pryce and Ward\cite{Pryce1947} and Snyder \emph{et al.}\cite{Snyder1948} determined the enhancement factor with different formalisms yielding strictly the same values.

Similarly, line $B_2$ of Table 1 of Bohm and Aharonov calculates the parallel and perpendicular scattering probabilities for orthogonally polarised non-entangled $\gamma$ as:
\begin{equation}
    P_\parallel(\theta) = 2\gamma^2-4\gamma{}sin^2\theta+sin^4\theta 
    \quad\mathrm{and}\quad 
    P_\perp(\theta) = 2\gamma^2-4\gamma{}sin^2\theta+3sin^4\theta.
    ~\footnote{Note that a typographical error in Bohm and Aharonov for the perpendicular term of linearly polarised photons (reported $2\gamma^2-4\gamma{}^2sin^2\theta+3sin^4\theta$) has here been corrected.}
\end{equation}
The non-entangled analytical prediction of the enhancement is plotted as the red line in Fig. \ref{fig:enh}. The widely reported maximum of 2.85 at 81.7$^\circ$ is only achieved if the fully entangled treatment is employed. Neglecting the entanglement significantly reduces the maximum enhancement to 1.63.

In order to calculate the normalised scattering probability as a function of $\Delta\phi$ for both entangled and orthogonally polarised non-entangled $\gamma$, we combine equations \ref{eqn:cos2phi} and \ref{eqn:enh} and solve for $\Delta\phi=0^\circ$ and $\Delta\phi=90^\circ$ yielding:
\begin{equation}
    Enh(\theta) = \frac{B(\theta)-A(\theta)}{B(\theta)+A(\theta)}
\end{equation}
which rearranges to:
\begin{equation}
    \frac{A(\theta)}{B(\theta)} = \frac{1-Enh(\theta)}{1+Enh(\theta)}.
\end{equation}
Thus, if we integrate equation \ref{eqn:enh} to obtain the enhancement factor over our range of interest ($Enh_{\Delta\theta}$) and take the normalisation $1/B$, we can calculate the relative double-scattering probability as:
\begin{equation}
    \frac{P_{\Delta\theta}}{B_{\Delta\theta}} = \frac{1-Enh_{\Delta\theta}}{1+Enh_{\Delta\theta}}\cdot{}cos(2\Delta\phi) + 1
\end{equation}

\begin{figure}
 \centering
\includegraphics[width=0.8\textwidth]{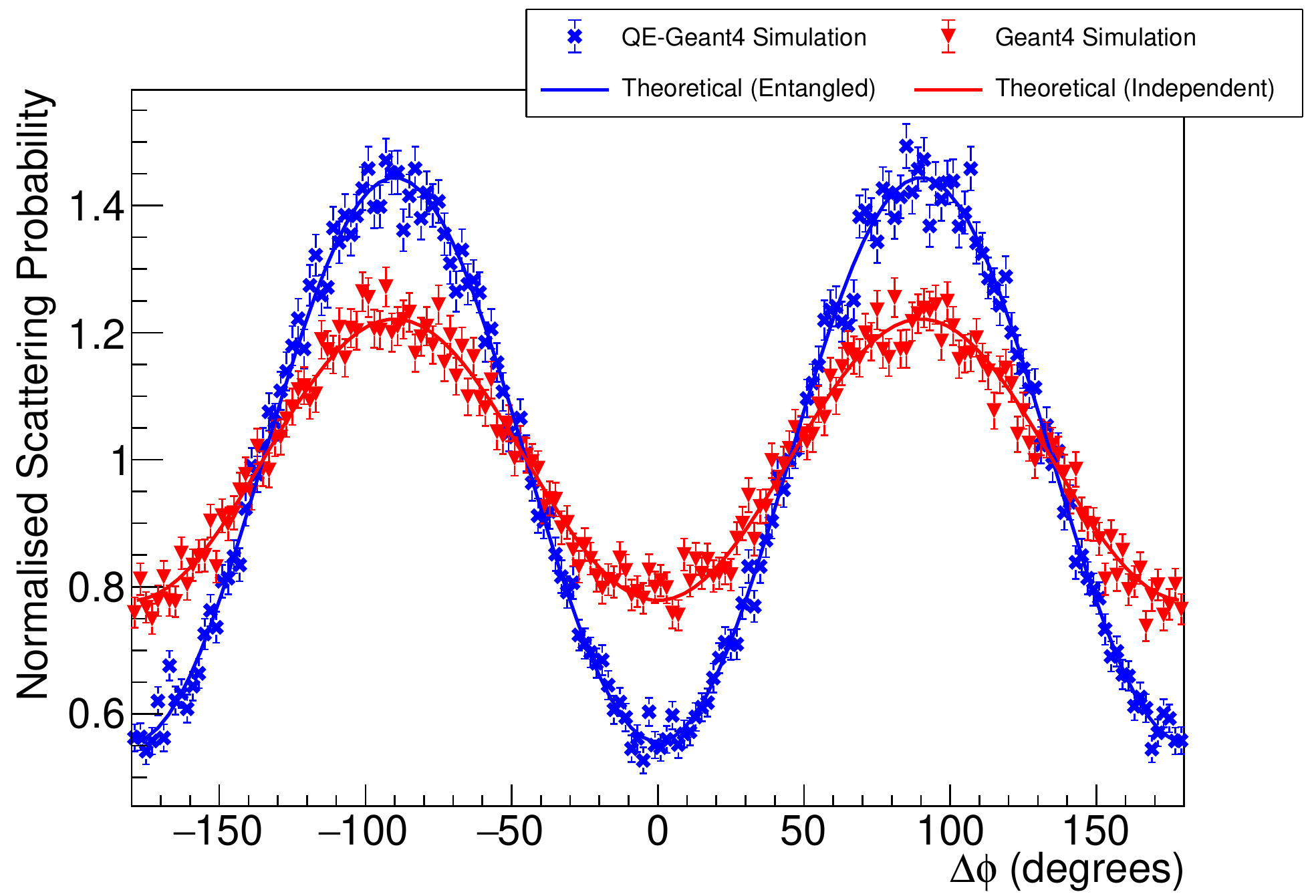} 
 \caption{\label{fig:simsVsTheory}\textbf{Check of the consistency of the predicted $\Delta\phi$ correlation from simulation and the underlying quantum theory.} Data points and curves show the normalised scattering probabilities as a function of the relative azimuthal scattering angle, $\Delta\phi$, between the two $\gamma$. The analytical theoretical predictions (solid lines) are shown for entangled (blue) and non-entangled orthogonally polarised (red) annihilation $\gamma$, calculated according to the formulae described in Table 1 of the publication of Bohm and Aharonov\cite{Bohm}. 
These are compared to QE-Geant4 simulation results for the entangled $\gamma$ (blue data points) and using standard Geant4 for the non-entangled $\gamma$ (red) scenarios. See text for details of the simulation.}
\end{figure}

The results of these calculations for our two cases are plotted as the solid lines in Fig. \ref{fig:simsVsTheory}. 
Satisfactory agreement is observed between the simulation and the theoretical predictions for each case, demonstrating that our implementation of quantum mechanical entanglement into Geant4 is consistent with theory, and that standard Geant4 does not fully account for this.

\subsection{Further statistical analysis of the agreement between experimental data and simulations}

\begin{figure}
 \centering
\includegraphics[width=0.8\textwidth]{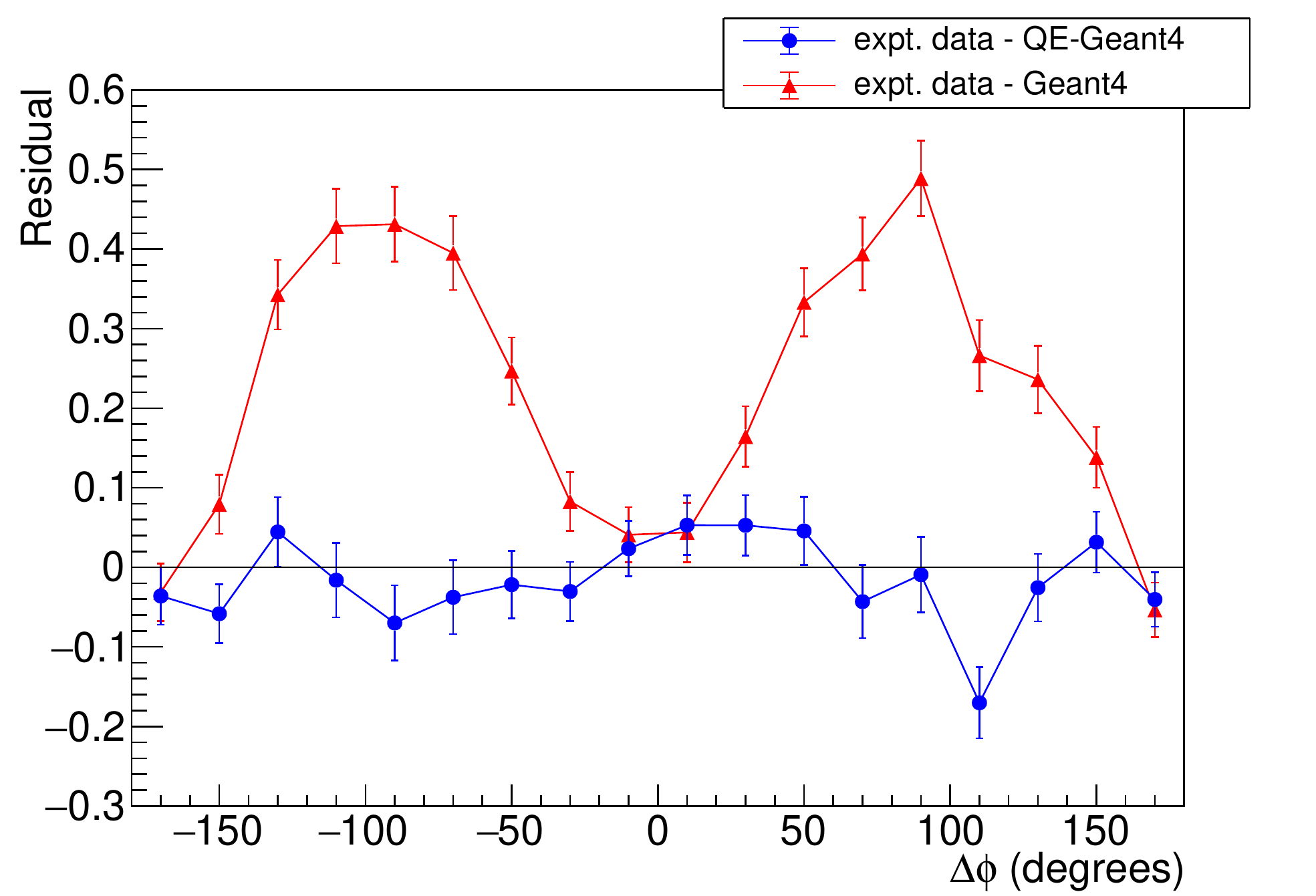} 
 \caption{\label{fig:residuals}\textbf{Residuals plot comparing the experimental data with simulations.} The bin-by-bin residuals (see text) obtained from a comparison of the experimental data with the simulation predictions (all analysis criteria as in Fig. \ref{fig:data}). The residuals for Geant4 (red) and QE-Geant4 (blue) simulations are both presented. The error bars are the quadrature sums of the statistical error of the experimental data and simulation for each bin.}
\end{figure}

In Fig. \ref{fig:residuals}, we present a bin-by-bin determination of the magnitude of the difference (the residual) between the experimental data and the simulation predictions presented in Fig. \ref{fig:data}. The QE-Geant4 simulation (blue data points) matches the shape of the experimental data well over all $\Delta\phi$, with residuals distributed about zero. The standard Geant4 simulation (red data points) however produces large residuals, failing to reproduce the amplitude of the experimentally observed $cos(2\Delta\phi)$ distribution.  
Furthermore, we have calculated $\chi^2/\nu$ values of 1.87  and 42.8 for the (expt. data - QE-Geant4) and the (expt. data - Geant4), respectively.

\subsection{The adopted PET imaging method}
A scanner consisting of an extension of our experimental PET-demonstrator apparatus was implemented in QE-Geant4 (Fig. \ref{fig:scanner}). To match the real detector position resolution of the experimental setup, the coordinates of the simulated energy deposits were smeared according to a Gaussian distribution with $\sigma = 0.8 $ mm$/\sqrt{12}$ . QE-PET event data, including $\theta$ and $\Delta\phi$, were stored in a list-mode file. They were sorted into three sets: the first with no condition on $|\Delta\phi|$, the second and the third contained only events within $0^\circ\le|\Delta\phi|\le20^\circ$ or $80^\circ\le|\Delta\phi|\le100^\circ$ windows, respectively. Image reconstruction was then performed, based on the methodology implemented in GAMOS (Geant4-based Architecture for Medicine-Oriented Simulations)\cite{GAMOS}. For each set, data were histogrammed into sinograms with the ``lm2pd" utility. Images were reconstructed with the implementation of the single-slice rebinning FBP (SSRB-FBP2D) algorithm\cite{Daube-Witherspoon1987}. A ramp filter was applied. Pixel size was set to $0.6 \times 0.6 $~mm$^{2}$. Images were processed with the NucMed plugin of ImageJ\cite{ImageJ}, as recommended by the GAMOS users guide.





\subsection{QE-PET method used to extract scatter and random contributions}

\begin{figure}
 \centering
\includegraphics[width=0.8\textwidth]{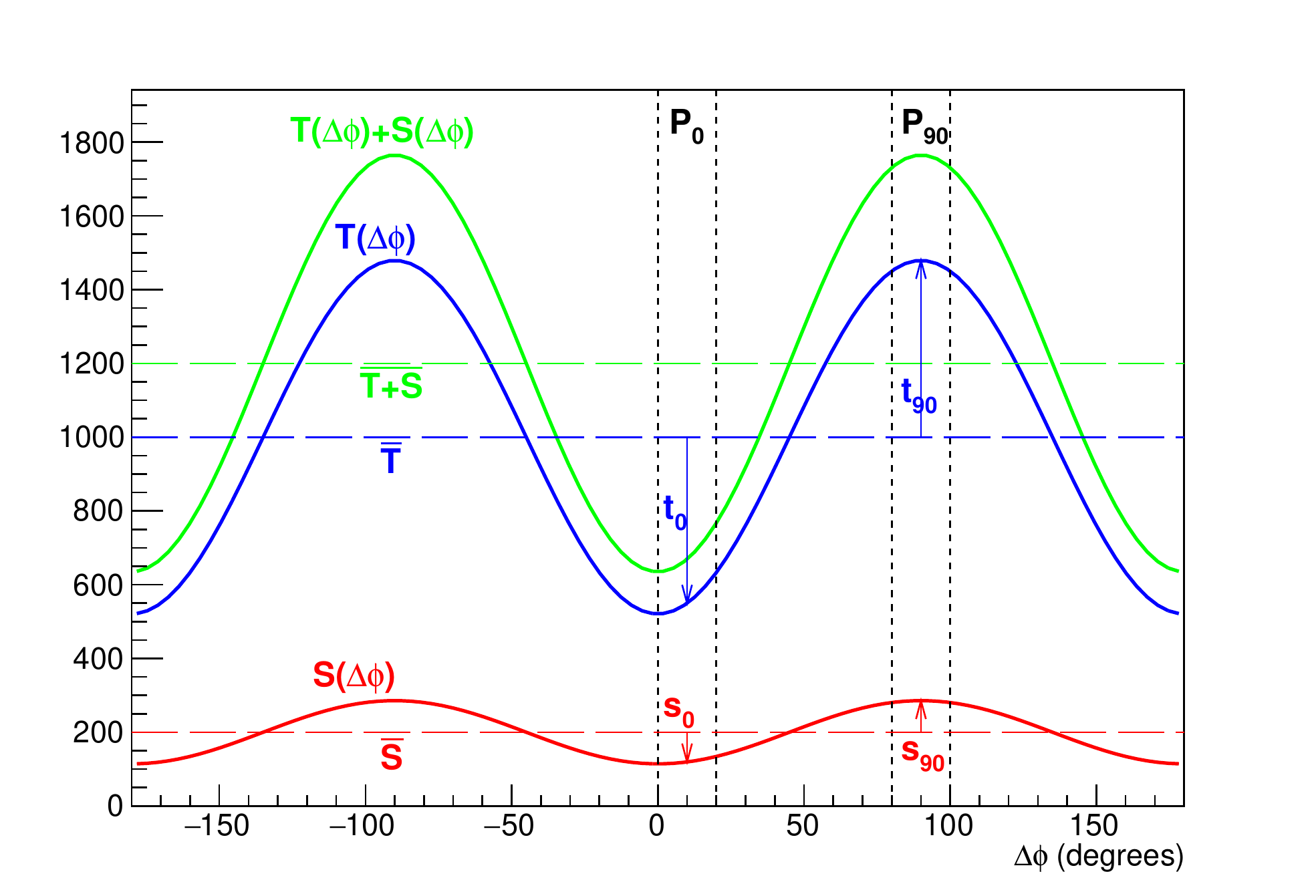}
 \caption{\label{fig:scatExtraction}\textbf{Illustration of the method to extract the scatter contribution.} Illustration of the different contributions to the total measured coincidences (green), from true coincidences (blue) and scatter events (red). The y-scale is arbitrary, and the relative number of scatter events and the amplitude of their $cos(2\Delta\phi)$ dependence have been exaggerated for illustrative purposes. Relevant features used to extract the scatter contribution from a combination of two $\Delta\phi$ slices are indicated (see text).}
\end{figure}

The coincidence count rate varies as a function of the relative azimuthal scattering angle $\Delta\phi$ (see for example, equation \ref{eqn:EKN} and  Fig.~\ref{fig:data}). For a given $\Delta\phi$, the total number of coincidences can be considered as the sum of true and scattered coincidences, i.e. $T(\Delta\phi) + S(\Delta\phi)$, as illustrated in Fig. \ref{fig:scatExtraction}. For a given range of $\Delta\phi$, the total number of coincidences can be expressed as a linear sum of the mean number of \emph{trues} and \emph{scatters}, weighted by appropriate scaling factors, e.g.
\begin{equation}
    P_{90} = t_{90}\cdot{}\bar{T} + s_{90}\cdot{}\bar{S}
    \quad\mathrm{and}\quad
    P_{0} = t_{0}\cdot{}\bar{T} + s_{0}\cdot{}\bar{S}
    \label{eqn:profiles}
\end{equation}
where $P_{90}$ and $P_{0}$ are the total number of coincidences in the $|\Delta\phi|$ windows centered around $90^\circ$ and $0^\circ$, respectively. $t_{90}=T(90^\circ)/\bar{T}$ and $s_{0}=S(0^\circ)/\bar{S}$ etc. These latter terms account for the small residual $\Delta\phi$ dependence for the scatter coincidences\footnote{From our fits to simulation, the residual enhancement for scatter coincidences is $Enh=1.207$, compared to 2.541 for the true coincidences.} arising from any remaining correlation in the polarisation planes of the detected $\gamma$\footnote{We should remark that the size of this residual enhancement is modelled assuming the expected collapse of the entangled state, as supported by the results presented in Fig 4. If future developments in our understanding of entanglement breaking at the MeV scale result in changes in the residual $\Delta\phi$ dependence then scaling factors can be extracted using the same method but with a simulation accounting for such effects}. Combining the two equations of \ref{eqn:profiles}, we can express the mean scatter contribution, $\bar{S}$, purely in terms of a linear combination of $P_{90}$ and $P_0$, i.e.
\begin{equation}
        \bar{S} = \frac{t_{90}\cdot{}P_{0}-t_{0}\cdot{}P_{90}}{t_{90}\cdot{}s_{0}-s_{90}\cdot{}t_{0}}.
        \label{eqn:extraction}
\end{equation}

The coefficients $s_0, s_{90}, t_0 $ and $t_{90}$ are extracted from fits to simulated QE-Geant4 data. If for $P_0$ and $P_{90}$ we use intensity profiles extracted from images generated with different cuts on $|\Delta\phi|$, we can use this methodology to extract a profile containing only the scatter contribution to the profile. A similar method can be applied to extract the \emph{true} coincidences. The process for extraction of random background is identical, except that the coefficients $s_0$ and $s_{90}$ are set to unity as the random coincidence count rate is independent of $\Delta\phi$.

\begin{figure}
 \centering
\includegraphics[width=0.95\textwidth]{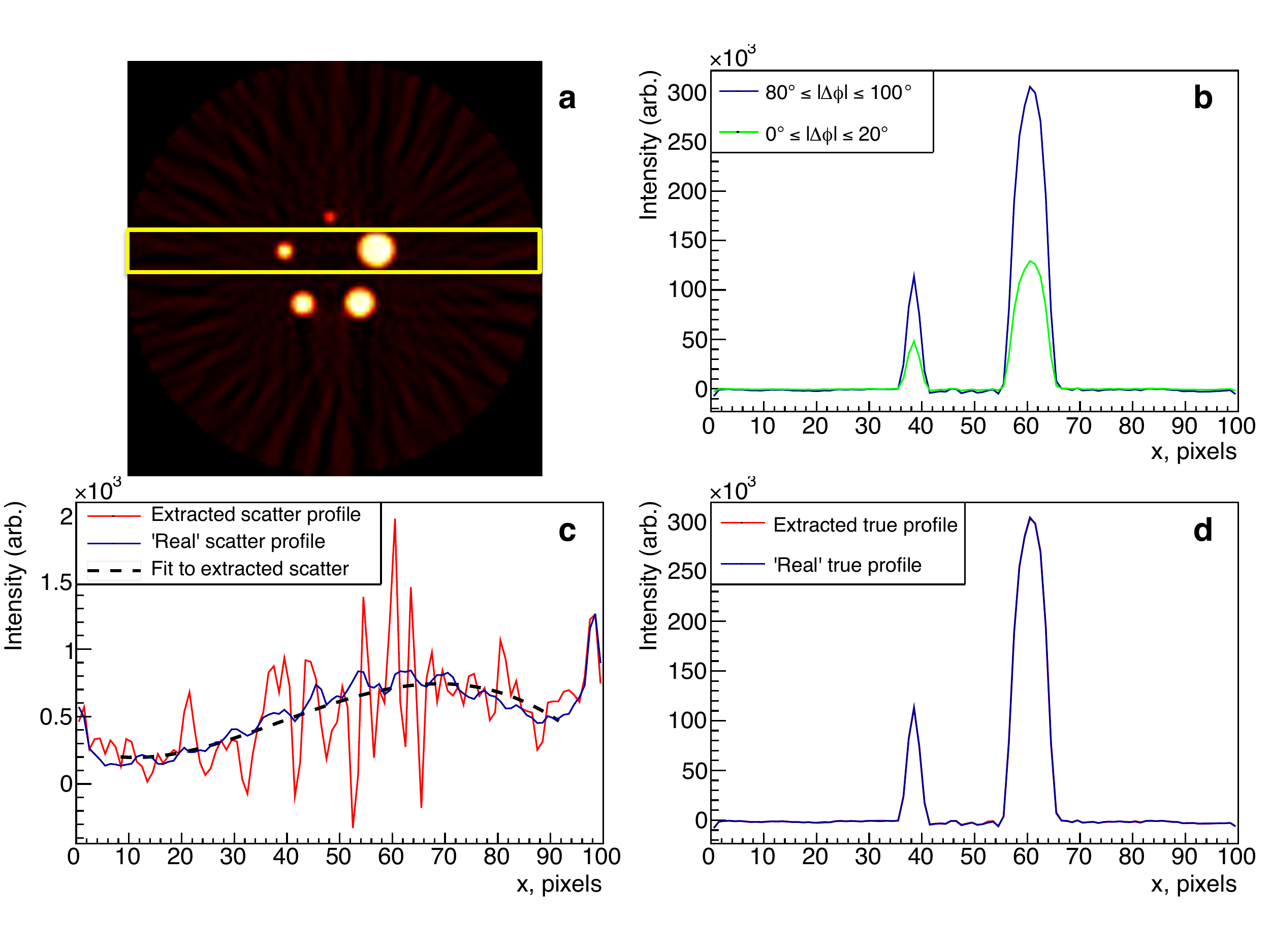}
 \caption{\label{fig:scatMiddle}\textbf{Extraction of true and scatter contributions from middle capillaries.}  (\textbf{a}) FBP two-dimensional PET image of the NEMA-NU4 phantom for true events with a scatter background. (\textbf{b}) Intensity profiles through the region indicated by the yellow rectangle for different $\Delta\phi$ cuts, i.e. $0^\circ\le|\Delta\phi|\le20^\circ$ (green), and $80^\circ\le|\Delta\phi|\le100^\circ$ (blue). (\textbf{c}) QE-PET profile for scatter background events extracted from a scaled subtraction of the two $\Delta\phi$ cut profiles (red line). The blue line shows the profile from the scatter events in isolation using the information from QE-Geant4. The dashed line is a $4^{th}$ order polynomial fit to the extracted scatter profile. (\textbf{d}) Profile extracted for true events with QE-PET (red line) compared to the profile of true events in isolation using QE-Geant4 (blue line).}
\end{figure}

Equation \ref{eqn:extraction} highlights that this method is independent of the relative numbers of true and scatter coincidences; this information is contained within the two $\Delta\phi$ slices. The only input required is the dependence of these components on $\Delta\phi$. For random events, this is constant while for scattered events there is a small residual $cos(2\Delta\phi)$ modulation. In the results, we adopted \emph{global} coefficients integrated over the full phantom. To test this assumption, we compared the $\Delta\phi$ dependence in different regions of the phantom and found that the amplitude remained consistent within statistical uncertainties ($\sim5\%$). Further checks were obtained from the extracted true and scatter contributions for profiles from different regions of the phantom, while using a single set of \emph{global} coefficients for the residual scatter profile. The method was found to perform satisfactorily in all cases studied. As an example, in Fig. \ref{fig:scatMiddle} we present profiles through the middle two capillaries. As was observed in Fig. \ref{fig:scatter}, the agreement between the extracted and ``actual'' true profiles is excellent. The underlying shape of the scatter profile is different to that seen in Fig. \ref{fig:scatter}c and in good agreement with the ``actual'' scatter profile.

\end{document}